\documentclass[12pt,preprint]{aastex}
\shorttitle{}
\shortauthors{}
\begin{document}
\title{Spectral Lags Obtained by CCF of Smoothed Lightcurves}
\author{Li, Zhao-sheng,  Chen Li, Wang, De-hua}
\affil{Department of Astronomy, Beijing Normal University,
Beijing, BJ 100875}
\email{chenli@bnu.edu.cn}

\begin{abstract}
We present a new technique to calculate the spectral lags of gamma-ray bursts
(GRBs). Unlike previous processing methods, we first smooth the light curves
 of gamma-ray bursts in high and low energy bands using the ``Loess" filter,
then, we directly define the spectral lags as such to maximize the
cross-correlation function (CCF) between two smoothed light curves.
This method is suitable for various shapes of CCF; it effectively
avoids the errors caused by manual selections for the
fitting function and fitting interval. Using the method, we have carefully measured
the spectral lags of individual pulses contained in BAT/Swift gamma-ray
bursts with known redshifts, and confirmed the anti-correlation between
the spectral lag and the isotropy luminosity. The distribution of
spectral lags can be well fitted by four Gaussian components,
with the centroids at 0.03 s, 0.09 s, 0.15 s, and 0.21 s, respectively. We find that
some spectral lags of the multi-peak GRBs seem to evolve with time.
\end{abstract}


\section{Introduction}
Observationally, the shape of a light curve of gamma-ray burst (GRB) is quite complex.
It contains one or several pulses characterized by a fast rise
followed by an exponential decay (FRED) profile \citep[e.g.,][]{fishman,feni99}.
The majority of light curves do not present periodic
variations. Light curves in different energy bands differ in many aspects,
such as the widths. The widths of the pulses in the higher energy bands
are usually smaller than those in the lower ones \citep{ferni95,nor05}. The time delay among
different energy photons is called spectral lag. As has been suspected,
the spectral lags of GRBs and their evolution are vital for probing the
physics of GRBs \citep{schaefer07}. Lots of statistical works have been done
\citep{band,nor2000,chen}. By using cross
correlation function, \citet{nor2000} estimated spectral lags of
six GRBs with known redshifts, concluding that the pulse peak luminosity
and the spectral lag in time $\tau_{\rm lag}$ are anticorrelated and can
be well fitted with a power law $L\thicksim\tau_{\rm{lag}}^{-1.14}$.
\citet{schaefer} explained the relation and also
demonstrated that the isotropic luminosity and the spectral lag should
meet $L\thicksim \tau_{\rm{lag}}^{-1}$. \citet{shen} interpreted the spectral lag as the
curvature effect of relativistic motion of GRB shells. Interestingly, \citet{chen}
examined the spectral lags of GRBs with multi-pulse, and found that there
seemed to be no apparent correlation between spectral lags and luminosity in
general for pulses within a given long GRB. Since Swift launched successfully,
many GRBs have been measured redshifts \citep{gehrels}. With larger samples, this anti-correlation
was checked carefully. The results show that the lag-luminosity correlation does
exist, but with a larger scatter \citep{schaefer07,xiao,ukwatta}.
The spectral lags in long and short GRBs are
quite different. Generally, short GRBs have nearly zero lags and long GRBs have
large positive lags (corresponding a temporal lead by higher energy  $\gamma$-ray photons) \citep{band}.
From this point of view, the spectral lags can be used as one of the
observational parameters to classify the GRBs.

The procedure for estimating spectral lags of GRBs using a cross correlation function (CCF) has been
widely adopted  \citep[e.g.,][]{link,ferni95,nor2000}.
The CCF of $x_{1}(t)$ and $x_{2}(t)$ for a GRB, where  $x_{1}(t)$
and $x_{2}(t)$ are the respective light
curves in two different $\gamma$-ray photon
energy bands, is simply defined as
\begin{equation}
\rm{CCF}_{\rm Std}(\tau; \nu_1, \nu_2)=\frac{\langle{\nu_1(t+\tau){\nu_2(t)}}\rangle}{\sigma_{\nu_1}\sigma_{\nu_2}},
\end{equation}
where $\nu_i(t)\equiv x_i(t)-\langle{x_i(t)}\rangle$
is the light curve of zero mean and
$\sigma_{\nu_i}=\langle{\nu{_i}^{2}}\rangle^{1/2}$ is the standard deviation from the mean.
The spectral lag $\tau_{\rm lag}$ is defined as such that it maximizes CCF($\tau$; $\nu_1$, $\nu_2$).
Because of the Poisson noise, $\tau_{\rm lag}$ has to be evaluated through fitting the maximum of
CCF with a polynomial function (Model I) or a Gaussian function with a linear term representing
a background (Model II). For a faint burst, the CCF often displays
asymmetry, or even multiple peaks. Therefore, using Model II to fit the CCF
maximum will introduce a systematic bias. Whereas it is difficult to 
determine the degree of polynomial to fit CCF with model I.
In both models, fitting interval of CCF will influence the result.

In order to reduce those man-made biases, we introduce a smooth technique.
In Sec.2, we smooth the light curves of different energy bands with ``Loess
method", and then calculate the CCF of two smoothed curves, finally, we
directly select the maximum point of CCF as the spectral lag.
The Monto-Carlo simulation is implemented to confirm the smooth factor $\alpha$. The algorithm
looks simple and reasonable. In Sec.3, we apply this procedure to 121
GRBs detected by Swift and compare the results with traditional algorithm.
The lag-luminosity correlation and lag-pulse width correlation are also
carefully analyzed in the third section. At last, we analyze the results
and give some brief discussions in section 4.

\section{Procedure of Analysis}

\subsection{Smooth the Light Curves First} \label{bozomath}
\begin{figure}
\epsscale{.80}
\plotone{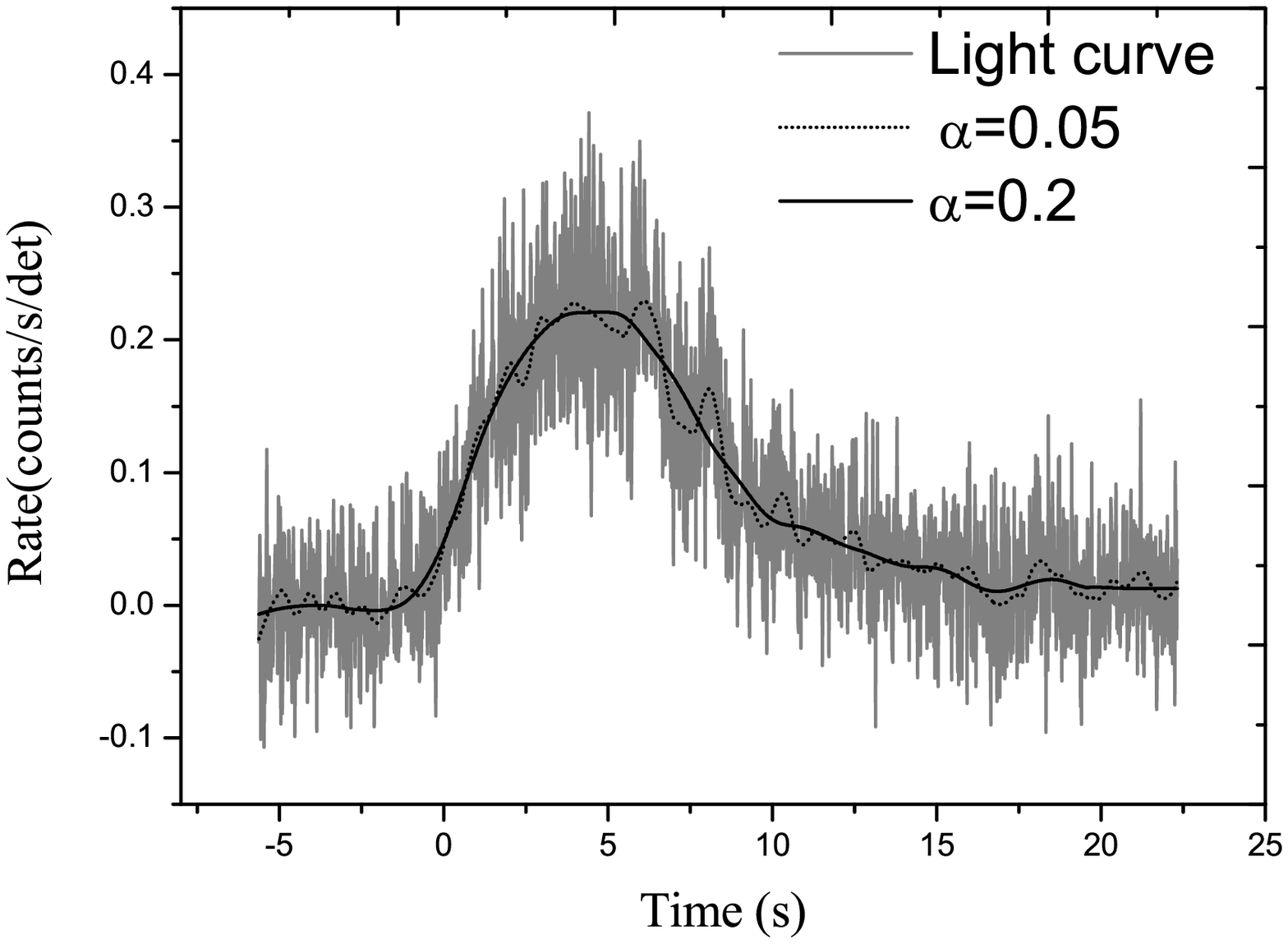}
\caption{Light curve of GRB 081222 with a temporal resolution 16ms.
The dotted and solid lines are the smoothed light curves
with $\alpha  = 0.05$ and $\alpha = 0.2$, respectively. \label{fig1}}
\end{figure}

We use a moving loess \citep{cleve,cleve88} filter
to smooth the GRB light curves. The Loess filter is
a local regression model, determined by
only one parameter: the smoothing factor,
$\alpha$.  $\alpha$ gives a percentage
($0\leq\alpha\leq1$), which means to take
$\alpha \times 100$ \% of the whole number of data
as the smooth-span. For example, supposing a light
curve contains 110 data points, $x_1,\cdots, x_{110}$,
taking  $\alpha = 0.1$ then a smoothed value
of $x_8$  is generated by a regression using linear
least squares with 11 data points $x_3, x_4,\cdots, x_{13}$
and a 2nd degree polynomial model. Obviously,
the $\alpha$ is smaller, the light curve is less smoothed, 
and vice versa. Especially,
when  $\alpha$ = 0, no smooth has been applied.
On the other hand, if  $\alpha$ is too large, the smoothed
light curve becomes very flat, i.e.  $\alpha = 1$,
the smooth-span is the interval of entire points.
Fig.~\ref{fig1} displays the smoothing results of GRB 081222 with
different $\alpha$. It is easy to see that $\alpha  = 0.2$ has stronger
smoothing effect than $\alpha= 0.05$.

In this paper, the smoothing procedure is as follows: Suppose we have a
single pulse GRB. First, we select an interval to cover the peak of
the GRB light curve (e.g., T90, the duration over which a burst emits
from 5\% of its total fluence to 95\%.) as time range for calculating spectral lag.
Then, we choose an appropriate smooth factor $\alpha$
which is determined by Monto-Carlo simulations, taking each $x_i$ as the
center of its smoothing span. We then use a second order 
polynomial model to fit all data points in the span, and replace $x_i$ by its fitted value.
Obviously, this is a moving average filter.

\citet{ukwatta} suggested that $\rm{CCF_{Std}}$ sometimes may not recover the artificial lag.
So, in the paper, we adopt the $\rm{CCF}$ defined by \citet{band},
\begin{equation}
{\rm{CCF}_{x,y}}(d) = \frac{{\sum\nolimits_{i = \max (1,1 - d)}^{\min (N,N - d)} {{x_i}{y_{i + d}}} }}{{\sqrt {\sum\nolimits_i {x_i^2} \sum\nolimits_i {y_i^2} } }}.
\end{equation}
Here, $x_i, y_i, i=0,\cdots, (N-1)$ denote the data of respective smoothed light
curves in two different energy bands.
The spectral lag is defined by $\tau_{\rm lag}=d \times t_b$, where $d$ is the maximum of CCF,
$t_{b}$ is the size of a time-bin.

\begin{figure}
\epsscale{.80}
\plotone{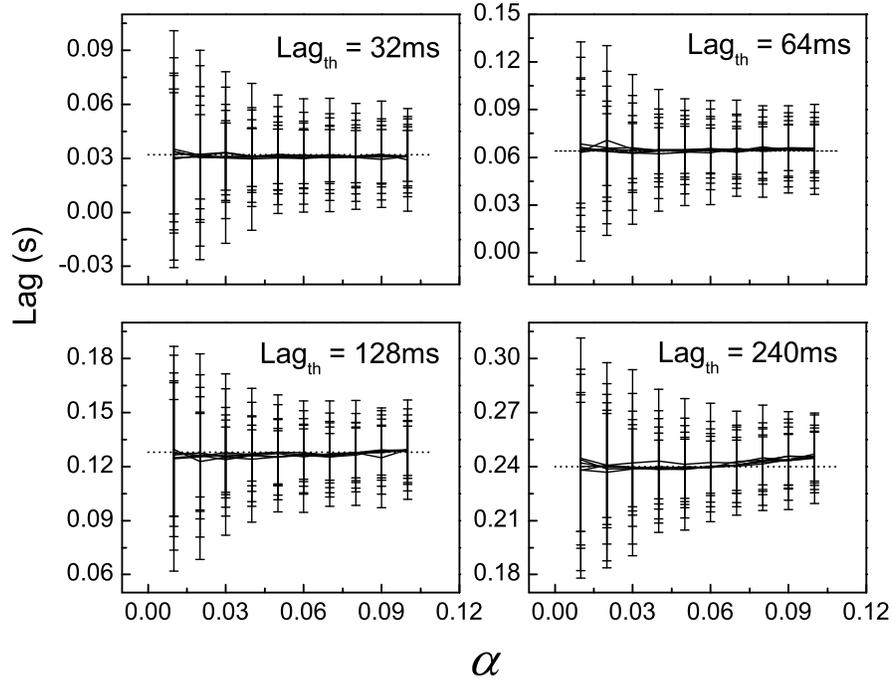}
\caption{Lag vs. $\alpha$.  The width ratio is 1.05
and the theoretical lags are 32 ms, 64 ms, 128 ms and 240 ms, respectively.
For each panel, the horizontal dotted line shows the theoretical lag.
The horizontal axis represents $\alpha$, and vertical axis represents lag.
The spectral lags and corresponding error bars of light curves are displayed,
with signal-to-noise changing from 5 to 10.  \label{fig2}}
\end{figure}

In order to determine a reasonable value of $\alpha$,
we calculated the lags between the simulated light-curves
of high and low energy bands. We selected the following
equation to model a pair of GRB light curves $F_h(t)$ (high energy band) and $F_l(t)$ (low energy band) \citep{abdo}:
\begin{equation}
F(t) = {C_0} + {p_0}\left\{ \begin{array}{l}
 0,t < {t_0} \\
 \frac{{t - {t_0}}}{\rho }\exp ( - \frac{{t - {t_0}}}{\rho }),t > {t_0}, \\
 \end{array} \right.
 \end{equation}
where $C_0$ is the background counts rate, $t_0$ is the trigger time,
$p_0$  and $\rho$  represent the amplitude and the width of light curves
respectively. Since our light curves were background-subtracted,
therefore $C_0=0$. The observational facts tell us that the light curves of higher
energy band have smaller $p_0$ and narrower width. In fact, the spectral lag
 obtained from the CCF has no relation to $p_0$. So we just need to consider
 the width ratios between two energy bands. In this paper, we choose
 width ratios as $1.05$. For a given width, we calculated
 the theoretical lags with equation (2). Then, we add a noise $X(t)$ on both
 $F_h(t)$ and $F_l(t)$, where $X(t)$ is a normally distributed random variable with
 expectation $0$ and standard deviation $\sigma$. Obviously, a larger $\sigma$
 corresponds to a lower signal-to-noise ratio (S/N). By adjusting the value of $\sigma$,
 we obtain the light curves with different S/N from $5$ to $10$. After smoothing,
 we achieved the simulation spectral lags of noisy light curves. We shift
 alpha from $0.01$ to $0.1$ (step=0.01) to examine which $\alpha$ can fit
 the theoretical lags best. Fig.~\ref{fig2} illustrates the simulating results of ``lag vs. $\alpha$".
 Each panel of Fig.~\ref{fig2} contains a dashed line and 6 solid curves. The dashed line
 shows the value of the theoretical lag, while the rest of the solid curves are the lags
 obtained from 6 different S/N, with higher S/N indicating shorter error-bars.
 All the curves reveal a similar trend: the curves are
 mildly flattened and shows little bit spread; all simulated lags are very close to
 the theoretical lags. When $\alpha$ ranges between $0.05$ and $0.1$, the largest relative errors
 between simulated and theoretical lags are less than $5\%$.
 In order to simplify calculations, our samples take $0.1$ and $0.05$ as the smooth factor,respectively.

As a comparison, we perform the CCF for a pair of original
light curves. In order to determine the maximum, we adopt
Model II to calculate lags between the light curves:
\begin{equation}
f(t) = a \cdot {\rm{exp}}( - {((t - b)/c)^2}) + d \cdot t + e,
\end{equation}
where $a,b,c,d$ are fitting parameters. The spectral lag is
${\tau _{{\rm{lag}}}} = {t_{\max }} \times {t_{\rm{b}}}$,
where $t_{\max }$ represent the time corresponding to the maximum of $f(t)$ ,
and $t_{\rm{b}}$ is the size of a time-bin.

\subsection{Uncertainty Estimation}

The Monte Carlo simulation is applied to estimate the
uncertainty of spectral lags. For a pair of given light
curves of high and low energy bands,

\begin{center}
$\left( \begin{array}{l}
 {x_1},{x_2}, \cdots ,{x_n} \\
 {e_1},{e_2}, \cdots ,{e_n} \\
 \end{array} \right) \rm{and} \left( \begin{array}{l}
 {y_1},{y_2}, \cdots ,{y_n} \\
 {\varepsilon _1},{\varepsilon _2}, \cdots ,{\varepsilon _n} \\
 \end{array} \right),$\\
\end{center}
where, $x_i$ and $y_i$ are count rates of high and low energy bands,
with measurement errors $e_i$ and $\epsilon_i$,  respectively. Based on the data,
we constructed a pair of simulated light curves
$( x'_1,x'_2, \cdots ,x'_n ) $ and $ (y'_1,y'_2, \cdots ,y'_n) $
by taking ${x'_i} \thicksim N({x_i},{e_i}^2)$ and ${y'_i} \thicksim N({y_i},{\varepsilon _i}^2),i = 1, \cdots ,n.$
Here,  ${x'_i}$ is a normally distributed random variable with expectation
$x_i$  and standard deviation $e_i$ , $i =1,\cdots,n$. The explanation is similar to ${y'_i}$.
In this work, we constructed 1000 (${x'_i}$ , ${y'_i}$) for each pair of observed $x_i$ and $y_i$.
We simulated 1000 light curves and derived the standard deviation of 1000 lags
as the uncertainty estimate of the lag.

\subsection{Determine the S/N and the duration of a GRB pulse}

\begin{figure}
\epsscale{.90}
\plotone{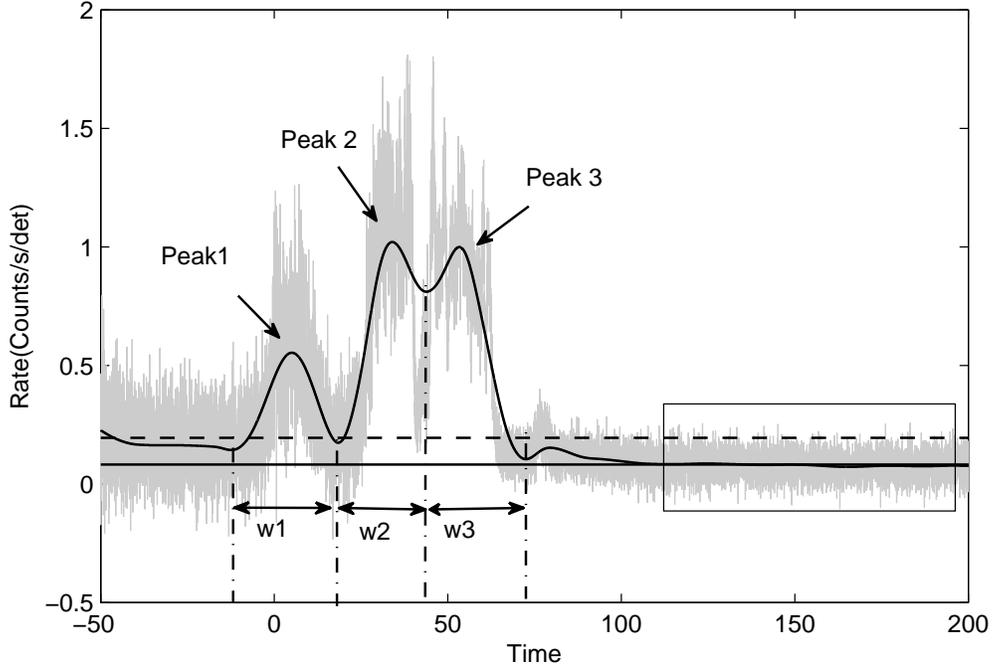}
\caption{Determine the pulse duration of GRB. The rectangle
region is considered as the GRB background, which standard
derivation denotes as $\sigma$. The height of horizontal
dashed line is minimum of smoothed light curve plus $\sigma$,
only the local maximum of solid line above the dashed line are regarded as pulse.
The vertical dot-dash lines represent the local minimums,
and the pulse duration is the interval between two dot-dash
lines ($w1$, $w2$, $w3$ are the duration for three pulses).
The height of horizontal solid line is the minimum of
smoothed light curve plus $0.1\sigma$. When the local maximum
misses the local minimum on one side or both sides, the
intersection points between the solid line and the
smoothed light curve are considered as start or end time of the pulse. \label{fig3}}
\end{figure}

The ratio between the maximum of a pulse and the standard deviation
of its background is defined as the S/N. Here the background is
an interval taken before or after the pulse region.
There are many works have been done to determine the
duration of GRB \citep{scar,hak}.
In this work, we still use the smoothing skill.
In Fig.~\ref{fig3}, we utilize GRB 061007 as an example
to represent the procedure of the duration of GRB pulses.
We smooth the GRB light curve for energy band 15-150 keV
with $\alpha=0.1$.
The smoothed light curve may contain a
few local maximum values. The lags which the pulse peaks
exceed $1\sigma$ background are returned. There are two
ways to determine the width of individual pulse. First,
when the pulse peak exists local minimums on both sides,
the pulse duration is the interval between the local minimums.
Second, when the pulse peak absents local minimum on one
side (or both sides), we drew a horizontal line which
height equals to the global minimum of smoothed light curve
with $0.1\sigma$ background added, and then the intersection
points between the horizontal line and the smoothed light
curve are considered as start or end point of the pulse.

\section{The GRBs sample and the results}
\subsection{Description of the GRBs sample}
Since the successful launch of Swift satellite in November 2004 \citep{gehrels},
over 600 GRBs have been detected. Swift is a multi-wavelength satellite
which can detect the gamma-ray transitional source and accurately locate
the source within less than 100 seconds. Swift observations have played
an inconceivably important role in GRB research.

The energy band of BAT/Swift is 15-350 keV. In practice, we only
choose photons between 15 and 150 keV because the BAT is transparent
to high energy photons over 150 keV. In previous works, GRBs light
curves were extracted by specifying GRBs positions which were detected
by BAT. Here, we use a slightly different method. The XRT can improve
GRB position in both accuracy and precision by using the UVOT to accurately
determine Swift position \citep{goad,evans}. So, we apply enhanced XRT
position 
to the procedure `batmaskwtevt' and `batbinevt' and extract
background subtracted 15-25 keV and
50-100 keV light curves with a time bin size of 16 ms for spectral lag calculating.

Obviously, when we apply this procedure to a low S/N GRB, it can produce unendurable error.
So, we select the GRBs with S/N larger than 5 in 15-25 keV energy range.

Our sample contains 121 long GRB detected by BAT/Swift from 2004 to 2010,
54 of them have measured redshifts.

\subsection{Results}
For a given spectral lag in advance, \citet{ukwatta}
simulated a group of light curves with the profile of a FRED pulse
superposed on a background of different noise levels and fitted
the peaks of CCF by a Gaussian function. The calculated lag was
consistent with the value given previously. Through calculation
we find that the CCF shows a symmetrical peak when the simulated
light curves have FRED-like pulse shapes. Obviously, Gaussian
curve is appropriate for fitting such maximum. The raw light curves,
however, are much more complicated than the simulated ones, it is
hard to get a good fitting with Gaussian function. As for our samples,
the shapes of the CCF can be roughly classified into 3 categories:
1) Gaussian-like profile; 2) Asymmetric peak and 3) multi-peaks.
Not difficult to imagine, it is easy to fit the first kind of
CCF with a Gaussian function but hard for the other two kinds.
We will compare the two methods with various shapes of CCF.

\begin{figure}
\epsscale{.90}
\plotone{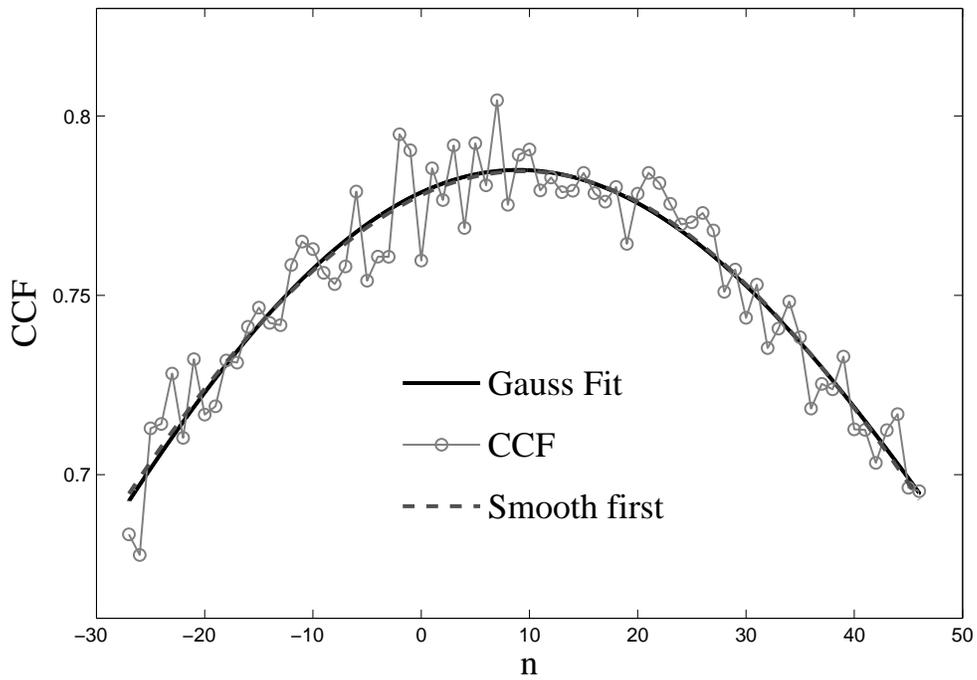}
\caption{A comparison between Gaussian fitting and loess filter methods for GRB 080413B. The black points represent the CCF between 15-25 keV and 50-100 keV light curves. The solid and dashed lines represent the results fitted by traditional Gaussian fitting or smooth method, respectively.\label{fig4}}
\end{figure}

We choose \object{GRB 080413B} and \object{GRB 071020} as examples. In Fig.~\ref{fig4},
the CCF between 15-25 keV and 50-100 keV of \object{GRB 080413B}
shows a Gaussian-like pulse. We fit the maximum of CCF with
a Gaussian function plus a linear function and obtain the
spectral lag and error, $\tau  = 0.14 \pm 0.02 $ s.
Then, using the smooth method, we obtain
$\tau  = 0.14 \pm 0.02 $ s. The results and figure
 show that either method can well model the spectral lag of \object{GRB 080413B}.

\begin{figure}
\epsscale{.90}
\plotone{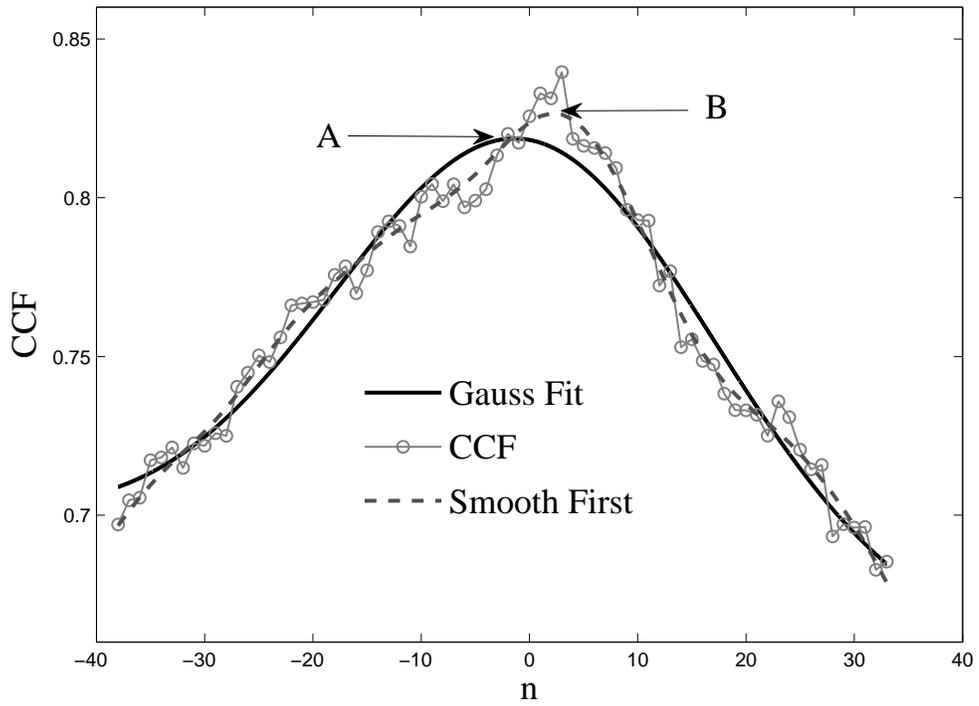}
\caption{A comparison between Gaussian fitting and loess filter methods for GRB 071020. A and B indicate the maximums of the CCF smoothed by Gaussian function and loess smooth method, respectively.\label{fig5}}
\end{figure}

In Fig.~\ref{fig5}, the CCF (circle) between 15-25 keV and 50-100 keV
of \object{GRB 071020} displays an asymmetry peak. There is an offset between
A and B. If we change the fitting interval, point A will move,
while the location of B is not related to the fitting interval.
Although using a smaller fitting interval that contains the maximum
of the CCF may yield similar lags, it will increase the lag uncertainty.
Fig.~\ref{fig5} shows the smooth method (dashed line) is
better for finding the maximum of CCF.

We list 121 spectral lags and pulse widths of GRBs in table~\ref{tb1-1} and table~\ref{tb1-2}.
For GRB with multiple pulses, we calculate each pulse of spectral lag. In the paper,
the spectral lags with smoothing factor $\alpha=0.1$ and $\alpha=0.05$ are utilized to analysis.

\subsection{The Results Analysis}

\subsubsection{Comparison between Gaussian Curve Fitting and Smooth Method}

From table~\ref{tb1-1} and table~\ref{tb1-2}, we notice that the correlation coefficient between
$\rm{lag}_{\rm{Gauss}}$ and $\rm{lag}_{\rm{loess}}$
is 0.9, which implies that the result of the two methods have high
correlation.  The $\rm{lag}_{\rm{Gauss}}-\rm{lag}_{\rm{loess}}$ relation
is fitted by a linear function
$\rm{lag}_{\rm{loess}}({\rm{s}}) = (0.75\pm0.05)\rm{lag}_{\rm{Gauss}}({\rm{s}}) + (0.005\pm0.03)$.
Hence, the $\rm{lag}_{\rm{Gauss}}$   is systematically larger
than $\rm{lag}_{\rm{loess}}$  by a factor of $4/3$.

\subsubsection{The Distribution of Spectral Lags With Smooth Method}
Distribution of the lags can be obtained as follows: Assuming
each spectral lag obeys a normal distribution with the mean
equals to itself and the standard deviation equals to its uncertainty.
In principle, the uncertainty of a lag should be larger than its
temporal resolution, 0.016 s. Therefore, if a simulated uncertainty is smaller
than 0.016 s, we set it to 0.016 s. In Fig.~\ref{fig7}, we add all probability
density function (PDF) together, and normalize the result.
As seen in Fig.~\ref{fig7}, the PDF of spectral lags has four components which
locate at $0.028\pm0.001$ s, $0.091\pm0.003$ s, $0.151\pm0.01$ s, and $0.21\pm 0.01$ s.
Obviously, most of GRBs have
positive spectral lags, which is consistent with the high energy
photons arriving earlier than those with low energy photons in long GRBs.

\subsubsection{The Relation between the Peak Isotropic Luminosity and the Lag of Primary Peak}
\citet{ukwatta} calculated spectral lags within the entire burst
region for the ``Gold sample" of GRBs detected by Swift,
confirming the correlation between the peak isotropic luminosity
and the lag of primary peak, albeit with a larger scatter in the relation. \citet{hakk} argued that
it is reasonable to calculate individual pulse spectral lag
instead of a burst range. From table~\ref{tb1-1} and table~\ref{tb1-2},
our results support that each pulse spectral lag of multi-pulse
GRBs has different delay time. We utilized spectrum and peak
flux data from \citet{bulter} and \citet{ukwatta},
and tested the lag-pulse isotropic luminosity relation.
Negative and zero lags were not shown in Fig.~\ref{fig8}. The best fit is
$\log L = (51.4\pm0.4) - (0.8\pm0.3)\log \rm{lag}/(1 + z)$.
The correlation coefficient R equals to -0.6, and the
slop $(0.8\pm0.3)$ cover the predicted slop of -1 by \citet{schaefer}.
Our results support the lag-luminosity relation.

\subsubsection{The Lag-pulse Duration Relation}

\citet{hakk} reported a correlation between
the spectral lag and pulse duration of GRBs with a high
correlation coefficient (R=0.97), i.e., the shorter the
duration of the pulse, the smaller the lag and the higher
the luminosity and vice versa.  We calculated the time
duration and pulse spectral lag of each pulse in the samples.
In Fig.~\ref{fig9}, the lag-duration relation in the rest frame of GRBs is still
establish but with a smaller correlation coefficient, R=0.6.

\subsubsection{The Evolution of the Lag with Time}
Most GRB light curves in the sample have multi-peak structure;
we show the lag corresponding to each peak in table~\ref{tb1-1} and table~\ref{tb1-2}.
We find that different pulses in one GRB generally have different
spectral lags, meaning that the lags evolve with time. Some GRBs
(\object{GRB 060927}, \object{GRB 061222A}, \object{GRB 080413A},
\object{GRB 080603B}, \object{GRB 090404}, \object{GRB 100615A})
even have different signs from different pulses, i.e. during a multi-peak burst,
one pulse has a positive lag, while the other may have a negative one.
It may be due to the time evolution of peak energy
which can produce negative lags \citep{peng,ukwatta11}.

\section{Conclusion}
\begin{figure}
\epsscale{.90}
\plotone{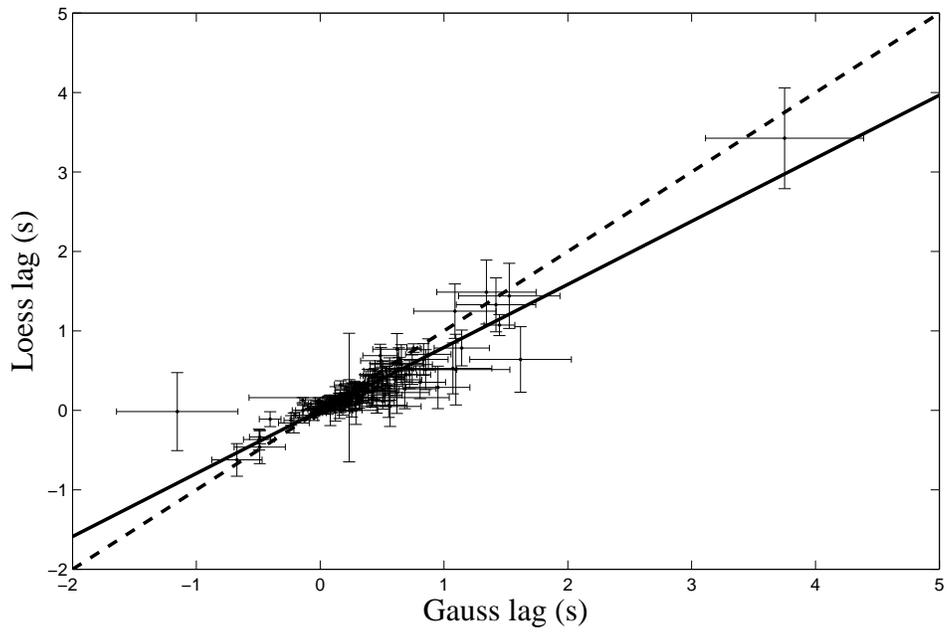}
\caption{$\rm{log}_{\rm Gauss}$ -$\rm{log}_{\rm loess}$ relation. The solid line shows the best fit between $\rm{log}_{\rm Gauss}$ and $\rm{log}_{\rm loess}$. The dashed line displays the diagonal. The $1\sigma$ simulation uncertainties are used for error bars.\label{fig6}}
\end{figure}

\begin{figure}
\epsscale{.90}
\plotone{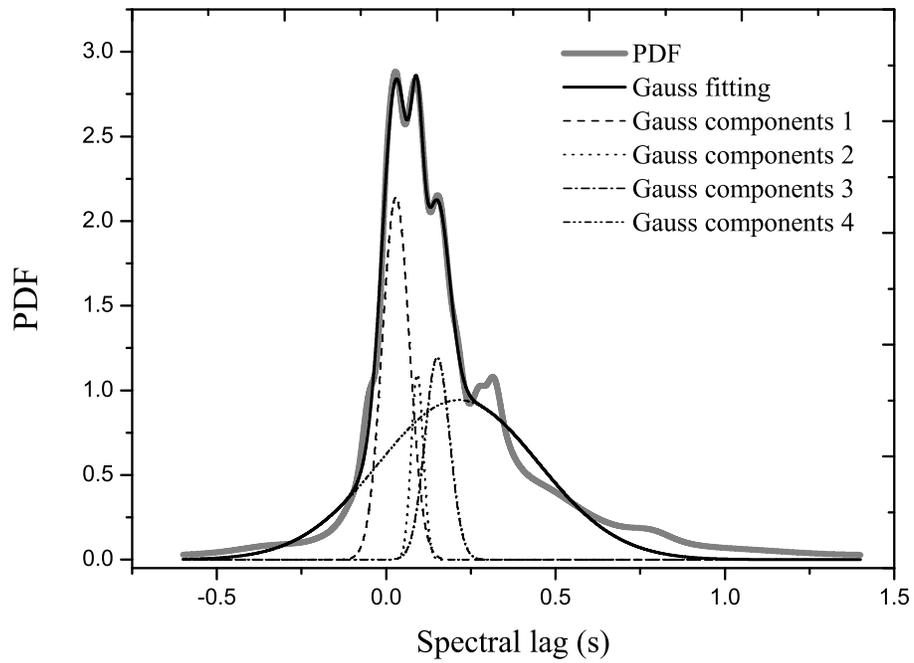}
\caption{The PDF of loess spectral lags. The PDF is fitted by four Gauss components
which locate $0.028\pm0.001$ s (dashed line), $0.091\pm0.003$ s (dotted line), $0.15\pm0.01$ s (dash-dot line), and $0.21\pm0.01$ s (dash-dot-dot line), respectively. \label{fig7}}
\end{figure}

\begin{figure}
\epsscale{.90}
\plotone{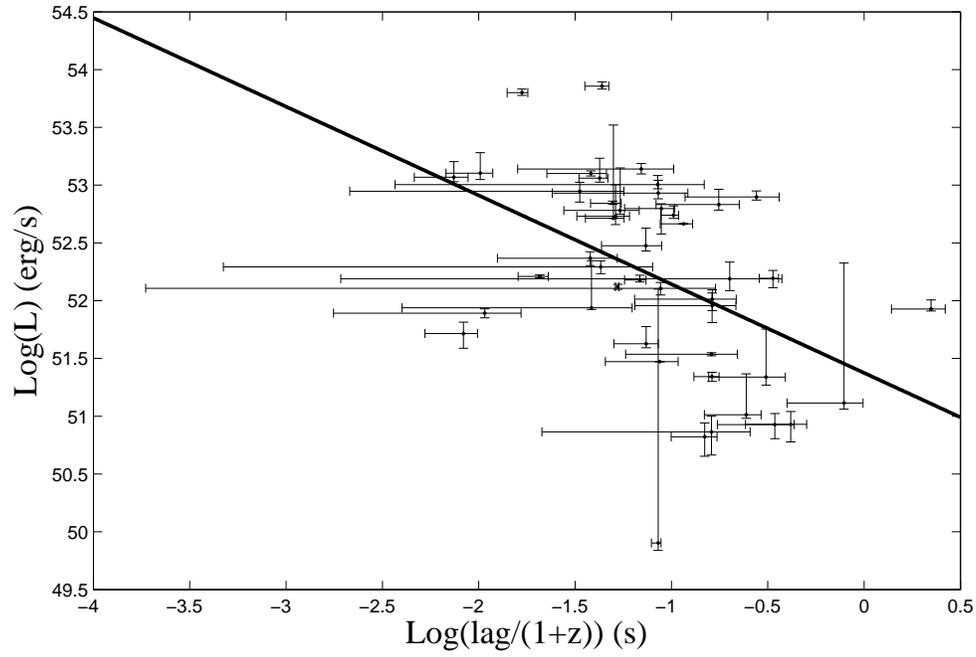}
\caption{The log-log relation between isotropic luminosity and spectral lag.
The factor $(1+z)^{-1}$  corrects for the time dilation effect.\label{fig8}}
\end{figure}

\begin{figure}
\epsscale{.90}
\plotone{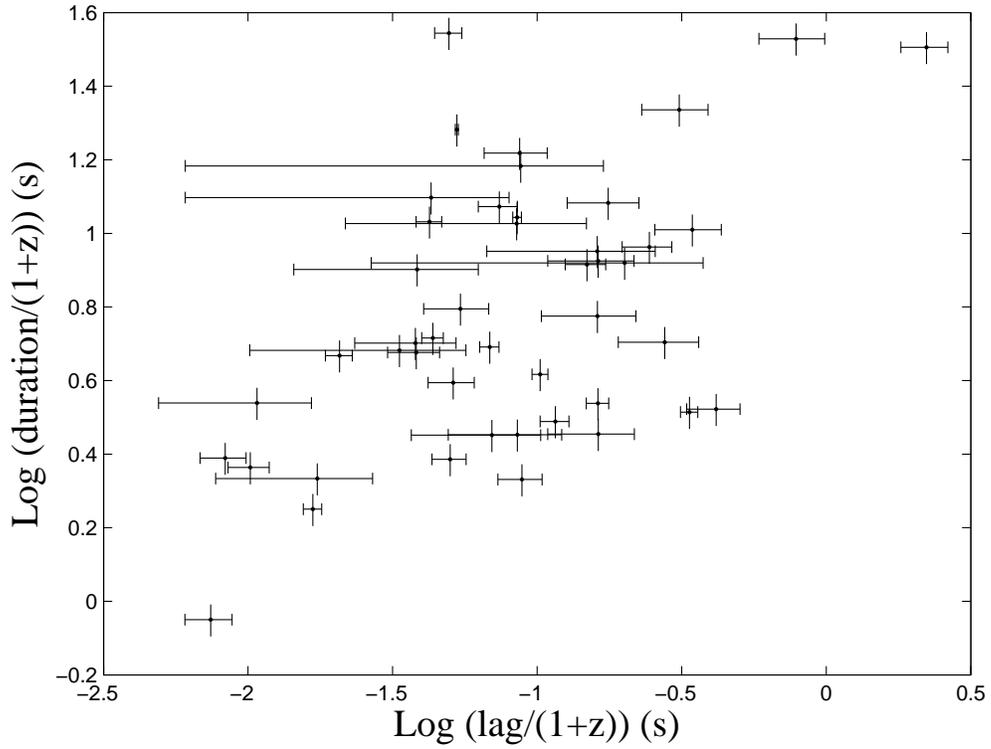}
\caption{Lag-pulse duration relation in the rest frame of GRBs.
Both axes are in units of second. The factor $(1+z)^{-1}$
corrects for the time dilation effect. Each duration error is set as $10\%$ of its value.\label{fig9}}
\end{figure}

In this work, we develop a new method to calculate the
spectral lags. Our method does not require the choose of fitting
function and intervals, thus avoiding the human selection effects in
traditional CCF fitting methods. The M-C simulation is utilized to determine the
smooth factor $\alpha$. The results show that our method obtains the introduced lags
appropriately as long as taking $\alpha $ between $0.05$ and $0.1$.
Using the method, we assign $\alpha=0.05$ and $\alpha=0.1$ to
calculate the spectral lags of GRBs detected by Swift BAT, respectively.
The Gaussian fitting and smoothing methods spectral lags list in
table~\ref{tb1-1} and table~\ref{tb1-2}. For two smoothing factors,
most of spectral lags cover each other well.
From Fig.~\ref{fig6}, we see spectral lags fitted by
Model II (i.e. Gauss+line) are strongly correlated
with the smooth method results, that demonstrates our method is
reasonable. It is worth noting that lags measured by our new
method are systematically smaller than those calculated by
traditional method. Figure 2 shows us that this is not caused by smoothing.
We also verify the isotropic luminosity-spectral
lag relation, which is consist with the work of \citet{nor2000} and \citet{ukwatta}.
By calculating multi-peak GRBs spectral lag, we find
lags evolve with time with a weak tendency. Finally,
\citet{hakk} reported the lag-pulse duration
relation with a extremely high correlation coefficient.
Our sample does not show this behavior.

\acknowledgments
We are grateful to the anonymous referee for a careful reading of this manuscript and for encouragement, constructive criticisms, valuable suggestions and helpful comments to improve the manuscript.
We thank to Yuan Tiantian for careful reading and polishing the manuscript. This work has been supported by the National Science Foundation of China (NSFC
10778716, NSFC 11173024 and NSFC 10773034),National Basic Research program of China 973
Projects (2009CB824800) and the Fundamental Research Funds for the Central Universities. This work made use of data supplied by the UK Swift Science Data Centre at the University of Leicester.

\clearpage

\clearpage

\clearpage

\clearpage

\clearpage

\clearpage

\clearpage

\clearpage

\clearpage

\clearpage

\begin{deluxetable}{lcccccc}
\tablewidth{0pt}
\tablecaption{The spectral lags of GRB with measured redshift.\label{tb1-1}}
\tablehead{
\colhead{GRB}  &\colhead{Peak No.}&\colhead{Start time} &\colhead {Stop time}
&\colhead{$\rm{lag}_{\rm{Gauss}}$}  & \colhead{$\rm{lag}_{\rm{Smooth}}$(s)}&
\colhead{$\rm{lag}_{\rm{Smooth}}$} \\
\colhead{ } &\colhead{ }&\colhead{}&\colhead{}&\colhead{ }&\colhead{$\alpha=0.1$}&\colhead{$\alpha=0.05$}\\
\colhead{ } &\colhead{ }&\colhead{(s)}&\colhead{(s)}&\colhead{(s)}&\colhead{(s)}&\colhead{(s)}\\
\colhead{(1)} &\colhead{(2)}&\colhead{(3)}&\colhead{(4)}&\colhead{(5)}&\colhead{(6)}&\colhead{(7)}
}
\startdata
050318	&        &     24.52    & 32.392    &    0.06	$\pm$     0.05	&    0	    $\pm$    0.05  &     0.11   $\pm$      0.06  \\
050401	&        &     22.968   & 33.992    &    0.44   $\pm$     0.1	&    0.27	$\pm$    0.13  &     0.14   $\pm$      0.05  \\
050416A	&        &     -1.648   & 3.856     &    0.49   $\pm$     0.1	&    0.69	$\pm$    0.14  &     0.64   $\pm$      0.1   \\
050603	&        &     -3.24    & 3.560     &    0.06   $\pm$	  0.02	&    0.06	$\pm$    0.02  &     0.05   $\pm$      0.02  \\
050922C	&        &     -3.624   & 4.152     &    0.21   $\pm$	  0.02	&    0.16	$\pm$    0.02  &     0.27   $\pm$      0.11  \\
051111	&        &     -8.88    & 12.304    &     1.1   $\pm$     0.4   &    0.51	$\pm$    0.44  &     0.48   $\pm$      0.2   \\
060206	&        &     -1.976   & 8.856     &    0.44   $\pm$	  0.07	&    0.45	$\pm$    0.08  &     0.43   $\pm$      0.2   \\
060210	&        &     -5.032   & 9.304     &    0.51   $\pm$	  0.2	&    0.43	$\pm$    0.2   &     0.34   $\pm$      0.1   \\
060418	&        &     23.104   & 34.848    &    0.08	$\pm$     0.2	&    0	    $\pm$     0.2  &     0.02   $\pm$      0.02  \\
060526	&        &     -5.040   & 16.192    &    0.25	$\pm$     0.06	&    0.16	$\pm$    0.06  &     0.21   $\pm$      0.08  \\
060614	&        &     -5.360   & 7.072     &    0.08	$\pm$     0.02	&    0.1	$\pm$    0.02  &     0.08   $\pm$      0.02  \\
060814	&   1    &     -0.696   & 29.704    &    0.33   $\pm$	  0.04	&    0.16	$\pm$    0.04  &     0.19   $\pm$      0.09  \\
        &   2    &     58.6     & 91.192    &    -0.49  $\pm$     0.1	&    -0.34	$\pm$    0.13  &     -0.6  $\pm$      0.4   \\
060904B	&        &     -11.488  & 25.424    &    0.64	$\pm$     0.1	&    0.53	$\pm$    0.14  &     0.58   $\pm$      0.2   \\
060912	&        &     -1.2     & 4.768     &    0.24	$\pm$     0.03	&    0.22	$\pm$    0.03  &     0.22   $\pm$      0.04  \\
060927	&   1    &     -1.864   & 3.896     &    0.06	$\pm$     0.02	&    0.05	$\pm$    0.02  &     0.08   $\pm$      0.02  \\
     	&   2    &     3.48     & 8.600     &    -0.02  $\pm$	  0.01	&    0	    $\pm$    0.02  &     0.21   $\pm$      0.1   \\
 	    &   3    &     13.16    & 29.064    &    -0.67  $\pm$	  0.2	&    -0.62	$\pm$    0.2   &    -0.74	 $\pm$      0.3   \\
061007	&   1    &     -4.296   & 19.768    &    0.52   $\pm$	  0.1	&    0.19	$\pm$    0.14  &     0.18	 $\pm$     0.05   \\
      	&   2    &     24.072   & 41.176    &    0.18   $\pm$	  0.06	&    0.16	$\pm$    0.06  &     0.11	 $\pm$     0.05   \\
     	&   3    &     41.592   & 65.928    &    0.17   $\pm$	  0.02	&    0.16	$\pm$    0.02  &     0.16	 $\pm$     0.01   \\
061121	&   1    &     -3.112   & 8.6       &    0.84   $\pm$	  0.2	&    0.64	$\pm$    0.2   &     0.82	 $\pm$     0.3    \\
     	&   2    &     58.712   & 82.312    &    0.02   $\pm$	  0.02	&    0.03	$\pm$    0.02  &     0.03	 $\pm$     0.02   \\
061201	&        &     -5.576   & 7.976     &    0.2    $\pm$	  0.03	&    0.19	$\pm$    0.03  &     0.08	 $\pm$     0.02   \\
070306	&        &     88.944   & 108.848   &    0.16   $\pm$	  0.06  &    0.1 	$\pm$    0.06  &     -0.08	 $\pm$     0.06   \\
070508	&        &     -5.712   & 29.12     &    0.09   $\pm$	  0.02	&    0.1 	$\pm$    0.02  &     0.1 	 $\pm$     0.02   \\
070521	&   1    &     7.3520   & 28.312    &    0.32   $\pm$     0.06	&    0.26	$\pm$    0.06  &     0.18	 $\pm$     0.04   \\
    	&   2    &     28.792   & 33.48     &    0.29   $\pm$	  0.03	&    0.14	$\pm$    0.03  &     0.11	 $\pm$     0.02   \\
    	&   3    &     33.416   & 40.232    &    0.08   $\pm$	  0.01	&    0.1 	$\pm$    0.01  &     0.14	 $\pm$     0.07   \\
070714B	&        &     -1.728   & 2.976     &    0.02   $\pm$	  0.02	&    0.02	$\pm$    0.02  &     0.02	 $\pm$     0.02   \\
070810A	&        &     -6.872   & 12.008    &    0.74   $\pm$	  0.2	&    0.51	$\pm$    0.2   &     0.5    $\pm$     0.3    \\
071003	&        &     -2.656   & 29.92     &    0.51   $\pm$	  0.09	&    0.11	$\pm$    0.1   &     0.11	 $\pm$     0.04   \\
071010B	&        &     -2.264   & 20.76     &    0.19   $\pm$	  0.02	&    0.14	$\pm$    0.02  &     0.14	 $\pm$     0.05   \\
071020	&        &     -3.808   & 3.456     &    -0.02  $\pm$     0.02	&    0.03	$\pm$    0.02  &     0.03	 $\pm$     0.02   \\
071117	&        &     -1.392   & 6.224     &    0.75	$\pm$     0.05	&    0.78	$\pm$    0.05  &     0.78	 $\pm$     0.06   \\
080210	&        &     -6.928   & 23.712     &    0.64	$\pm$     0.2	&    0.59	$\pm$    0.2   &     0.53	 $\pm$     0.3    \\
080319B	&        &     -5.792   & 62.032    &    0.2	$\pm$     0.02	&    0.1 	$\pm$    0.01  &     0.35	 $\pm$     0.02   \\
080319C	&        &     -2.784   & 15.6     &    0.23	$\pm$     0.04	&    0.16	$\pm$    0.04  &     0.19	 $\pm$     0.08   \\
080330	&        &     -2.432   & 15.264   &    0.16	$\pm$     0.05	&    0.32	$\pm$    0.05  &     0.03	 $\pm$     0.02   \\
080411	&   1    &     14.960   & 23.36     &    0.3	$\pm$     0.02	&    0.21   $\pm$     0.01 &     0.2 	 $\pm$     0.02   \\
	   &   2    &     38.720   & 49.04     &    0.2	$\pm$     0.02   &    0.18	$\pm$    0.02  &     0.18	 $\pm$     0.02   \\
    	&   3    &     52.608   & 62.624    &    0.29	$\pm$     0.2 	&    0	    $\pm$     0.2  &     0	    $\pm$     0.2    \\
    	&   4    &     62.512   & 72.192    &    0.23	$\pm$     0.8	&    0.16	$\pm$    0.1   &     0.11 	 $\pm$     0.1    \\
080413A	&   1    &     -2.888   & 10.6    &    0.2	    $\pm$     0.03	&    0.18	$\pm$    0.03  &     0.18	 $\pm$     0.05   \\
    	&   2    &     12.104   & 33.72   &    -0.05    $\pm$     0.03	&    -0.1	$\pm$    0.03  &     -0.11	 $\pm$     0.06   \\
    	&   3    &     35.032   & 58.264  &    0.66	    $\pm$     0.3	&    0.58	$\pm$    0.25  &     -0.05	 $\pm$     0.1    \\
080413B	&        &     -2.576   & 7.744   &    0.14	    $\pm$     0.02   &    0.14	$\pm$    0.02  &     0.13	 $\pm$     0.02   \\
080430	&        &     -1.288   & 14.936   &    0.42	$\pm$     0.08	&    0.43	$\pm$    0.08  &     0.51	 $\pm$     0.2    \\
080603B	&   1    &     -1.176   & 6.776    &    -0.2	$\pm$     0.03	&    -0.06	$\pm$    0.04  &     -0.03	 $\pm$     0.02   \\
    	&   2    &     6.6      & 19.688   &    0.38	$\pm$     0.06	&    0.27	$\pm$    0.06  &     0.37	 $\pm$     0.1    \\
    	&   3    &     38.872   & 74.04    &    0.29	$\pm$     0.08	&    0.27	$\pm$    0.08  &     0.19	 $\pm$     0.08   \\
080605	&        &     -6.84    & 21.544     &    0.12	$\pm$     0.02	&    0.11	$\pm$    0.02  &     0.11	 $\pm$     0.01   \\
080607	&        &     -7.296   & 13.664    &    0.16	$\pm$     0.02	&    0.18	$\pm$    0.02  &     0.14	 $\pm$     0.01   \\
080707	&        &     -11.864  & 10.952    &    0.62	$\pm$     0.2	&    0.77	$\pm$    0.2   &     1.12	 $\pm$     0.5    \\
080810	&        &     -17.824  & 43.008    &    -0.03  $\pm$     0.02	&    0	    $\pm$    0.02  &     0	    $\pm$     0.02   \\
080916A	&        &     -7.568   & 49.52     &    1.42	$\pm$     0.3	&    1.33	$\pm$    0.3   &     1.25	 $\pm$     0.4    \\
080928	&        &     195.432  & 225.016  &    -0.08   $\pm$     0.05	&    0.08	$\pm$    0.05  &     0.08	 $\pm$     0.03   \\
081203A	&        &     -0.224   & 47.056    &    0.68	$\pm$     0.2	&    0.3 	$\pm$    0.2   &     0.3 	 $\pm$     0.1    \\
081222	&        &     -1.888   & 16.000    &    0.31	$\pm$     0.02	&    0.14	$\pm$    0.03  &     0.14	 $\pm$     0.02   \\
090423	&        &     -12.84   & 30.936   &    0.39	$\pm$     0.2	&    0.30	$\pm$    0.2   &     0.24	 $\pm$     0.1    \\
090424	&   1    &     -1.784   & 5.4      &    0.01	$\pm$     0.02	&    0.03	$\pm$    0.02  &     0.03	 $\pm$     0.02   \\
    	&   2    &     5.656    & 26.616    &    0.08	$\pm$     0.08	&    0.11	$\pm$    0.08  &     0.08	 $\pm$     0.07   \\
090510	&        &     -1.888   & 6.4      &    0.03	$\pm$     0.02	&    0	    $\pm$     0.01 &     0	    $\pm$     0.02   \\
090618	&   1    &     -7.568   & 41.776   &    3.75	$\pm$     0.6	&    3.42	$\pm$    0.6   &     3.4 	 $\pm$     0.9    \\
     	&   2    &     51.2     & 72.224   &    0.46	$\pm$     0.03	&    0.3 	$\pm$    0.05  &     0.3 	 $\pm$     0.1    \\
    	&   3    &     72.112   & 99.52    &    0.001   $\pm$     0.02	&    0	    $\pm$     0.02 &     0	    $\pm$     0.02   \\
    	&   4    &     99.088   & 140.384  &    0.55    $\pm$     0.1 	&    0.3 	$\pm$    0.1   &     0.3 	 $\pm$     0.2    \\
090715B	 &  1    &     -13.952  & 34.432   &    0.87    $\pm$     0.2	&    0.7	$\pm$    0.2   &     0.4	 $\pm$     0.2    \\
    	&   2    &     48.56    & 89.2     &    0.61	$\pm$     0.2	&    0.38	$\pm$    0.2   &     0.53	 $\pm$     0.3    \\
091018	&        &     -1.112   & 5.688    &    0.35	$\pm$     0.03	&    0.32	$\pm$    0.03  &     0.32	 $\pm$     0.05   \\
091127	&   1    &     -1.2     & 3.952    &    0.02	$\pm$     0.01	&    0.02	$\pm$    0.02  &     0.02	 $\pm$     0.02   \\
    	 &  2    &     4.928    & 11.856   &    0.04	$\pm$     0.03	&    0	    $\pm$     0.02 &     0.03	 $\pm$     0.05   \\
091208B	 &  1    &     -1.088   & 4.784    &    0.36	$\pm$     0.1	&    0.34	$\pm$    0.1   &     0.34	 $\pm$     0.1    \\
    	 &  2    &     5.328    & 17.904   &    0.07	$\pm$     0.02	&    0.06	$\pm$    0.02  &     0.03	 $\pm$     0.02   \\
100316B	 &       &     -6.696   & 12.792   &    0.81	$\pm$     0.2	&    0.35	$\pm$    0.2   &     0.34   $\pm$      0.1   \\

\enddata

\tablecomments{col.(1): The GRB trigger number.
Col.(2): The peak number of GRB.
Col.(3): The start time of individual pulse relative to the GRB trigger time.
Col.(4): The stop time of individual pulse relative to the GRB trigger time.
Col.(5): The spectral lags plus errors between 15-25 keV
and 50-100 keV by fitting with Gaussian plus linear equation.
Col.(6):The spectral lags plus errors calculated
by smooth method with $\alpha=0.1$.
Col.(7):The spectral lags plus errors calculated
by smooth method with $\alpha=0.05$.}
\end{deluxetable}

\clearpage
\begin{deluxetable}{lcccccc}
\tablewidth{0pt}
\tablecaption{The spectral lags of GRB with unknown redshift.\label{tb1-2}}
\tablehead{
\colhead{GRB}  &\colhead{Peak No.}&\colhead{Start time} &\colhead {Stop time}
&\colhead{$\rm{lag}_{\rm{Gauss}}$}  & \colhead{$\rm{lag}_{\rm{Smooth}}$(s)}&
\colhead{$\rm{lag}_{\rm{Smooth}}$} \\
\colhead{ } &\colhead{ }&\colhead{}&\colhead{}&\colhead{ }&\colhead{$\alpha=0.1$}&\colhead{$\alpha=0.05$}\\
\colhead{ } &\colhead{ }&\colhead{(s)}&\colhead{(s)}&\colhead{(s)}&\colhead{(s)}&\colhead{(s)}\\
\colhead{(1)} &\colhead{(2)}&\colhead{(3)}&\colhead{(4)}&\colhead{(5)}&\colhead{(6)}&\colhead{(7)}
}
\startdata
041220	 &        &      -2.608  &  8.096   & 	0.21  $\pm$ 	0.05 &	0.08     $\pm$ 	0.05 &	0.18  $\pm$	0.08 \\
041224	 &        &      16.032  &  41.568  & 	0.49  $\pm$ 	0.2  &	0.62     $\pm$ 	0.2  &	0.77  $\pm$	0.4  \\
050124	 &        &      -6.264  &  6.904   & 	0.03  $\pm$ 	0.02 &	0.08     $\pm$ 	0.02 &	0.05  $\pm$	0.02 \\
050219B	 &    1   &      -6.376  &  5.432   & 	0.23  $\pm$ 	0.06 &	0.05     $\pm$ 	0.06 &	0.03  $\pm$	0.02 \\
     	 &    2   &      -5.976  &  15.896  & 	0.32  $\pm$ 	0.05 &	0.29     $\pm$ 	0.05 &	0.10  $\pm$	0.03 \\
050326	 &    1   &      -2.736  &  13.392  & 	0.03  $\pm$ 	0.02 &	0.02     $\pm$ 	0.02 &	0.02  $\pm$	0.02 \\
     	 &    2   &       15.52  &  30.560  & 	0.09  $\pm$ 	0.02 &	0.06     $\pm$ 	0.02 &	0.06  $\pm$	0.02 \\
050418	 &        &      -25.416 &  31.000  & 	0.44  $\pm$ 	0.09 &	0.50     $\pm$ 	0.1  &	0.42  $\pm$	0.2  \\
050509A	 &        &      -8.664  &  8.680   & 	0.06  $\pm$ 	0.03 &	0.14     $\pm$ 	0.04 &	0.06  $\pm$ 0.02 \\
050701	 &        &       2.968  &  13.384  & 	0.21  $\pm$ 	0.05 &	0.14     $\pm$ 	0.06 &	0.29  $\pm$	0.2  \\
050717	 &    1   &      -1.136  &  16.672  & 	0.01  $\pm$ 	0.02 &	0.05     $\pm$ 	0.02 &	0.03  $\pm$	0.02 \\
     	 &    2   &      16.368  &  39.408  & 	0.46  $\pm$ 	0.2  &	0.13     $\pm$ 	0.2  &	0.10  $\pm$	0.03 \\
050801	 &        &      -2.464  &  9.9680  & 	0.29  $\pm$ 	0.08 &	0.10     $\pm$ 	0.09 &	0.00  $\pm$	0.02 \\
050820B	 &    1   &      -1.608  &  5.3840  &   0.42  $\pm$ 	0.2  &	0.24     $\pm$ 	0.2  &	0.11  $\pm$	0.03 \\
     	 &    2   &      5.2240  &  17.688  &  -0.40  $\pm$ 	0.09 &	-0.11    $\pm$ 	0.09 &	-0.21 $\pm$ 0.07 \\
051016B	 &    1   &      -0.584  &  2.056   &  0.26   $\pm$ 	0.08 &	0.14     $\pm$ 	0.07 &	0.13  $\pm$	0.07 \\
     	 &    2   &       2.056  &  7.480   &  -0.49  $\pm$ 	0.2  &	-0.46    $\pm$ 	0.2  &	-1.07 $\pm$ 0.6  \\
060105	 &    1   &      -12.00  &  18.856  & 	0.43  $\pm$ 	0.09 &	0.16     $\pm$ 	0.1  &	0.16  $\pm$	0.05 \\
     	 &    2   &      19.256  &  43.368  & 	0.55  $\pm$ 	0.2  &	0.06     $\pm$ 	0.2  &	0.06  $\pm$	0.02 \\
060110	 &        &      -2.104  &  14.248  & 	1.09  $\pm$ 	0.3  &	1.25     $\pm$ 	0.3  &	1.34  $\pm$	0.6  \\
060111A	 &        &      -7.456  &  18.608  & 	1.53  $\pm$ 	0.4  &	1.44     $\pm$ 	0.4  &	1.95  $\pm$	0.5  \\
060117	 &    1   &      -2.776  &  5.176   &   0.12  $\pm$ 	0.02 &	0.08     $\pm$ 	0.02 &	0.08  $\pm$	0.02 \\
     	 &    2   &      5.5760  &  18.936  & 	0.08  $\pm$ 	0.02 &	0.03     $\pm$ 	0.02 &	0.03  $\pm$	0.02 \\
060223B	 &        &      -7.840  &  8.400   &  -1.15  $\pm$ 	0.5  &	-0.02    $\pm$ 	0.5  &	0.03  $\pm$	0.02 \\
060306	 &        &      -3.584  &  8.320   & 	0.18  $\pm$ 	0.02 &	0.08     $\pm$ 	0.03 &	0     $\pm$	0.02 \\
060313	 &        &      -3.024  &  2.048   &   0     $\pm$	   0.02  &	0   	 $\pm$	0.02 &	0     $\pm$	0.02 \\
060428A	 &        &      -3.136  &  11.904  & 	0.18  $\pm$ 	0.06 &	0.00     $\pm$ 	0.06 &	0.05  $\pm$ 0.02 \\
060708	 &        &      -1.976  &  10.392  & 	0.95  $\pm$ 	0.3  &	0.29     $\pm$ 	0.3  &	0.29  $\pm$	0.1  \\
060719	 &    1   &      -0.888  &  13.352  & 	0.85  $\pm$ 	0.3  &	0.51     $\pm$ 	0.3  &	0.5   $\pm$	0.2  \\
    	 &    2   &      40.136  &  60.168  & 	0.51  $\pm$ 	0.1  &	0.26     $\pm$  0.1  &	0.32  $\pm$	0.1  \\
060813	 &        &      -0.888  &  11.192  & 	-0.01 $\pm$  	0.02 &	0.03 	 $\pm$ 0.02  &	0.03  $\pm$	0.02 \\
060825	 &        &      -4.272  &  8.016   & 	0.54  $\pm$ 	0.1  &	0.43 	 $\pm$ 0.1   &	0.45  $\pm$	0.2  \\
060904A	 &        &      41.336  &  81.256  & 	-0.02 $\pm$  	0.02 &	0.1 	 $\pm$ 0.02  &	0.1   $\pm$	0.04 \\
061004	 &        &      -0.008  &  11.512  & 	-0.01 $\pm$  	0.03 &	0.13 	 $\pm$ 0.03  &	0.13  $\pm$	0.04 \\
061021	 &        &           0  &  10.880  & 	0.03  $\pm$ 	0.02 &	0    	 $\pm$ 0.02  &	0     $\pm$	0.02 \\
061126	 &        &      2.488   &  14.920  & 	0.14  $\pm$ 	0.03 &	0.13 	 $\pm$ 0.03  &	0.24  $\pm$	0.1  \\
061202	 &        &      70.648  &   95.976 & 	-0.14 $\pm$  	0.06 &	0   	 $\pm$ 0.06  &	0     $\pm$	0.07 \\
061222A	 &    1   &       23.184 &   39.984 & 	-0.23 $\pm$  	0.06 &	-0.13 	 $\pm$ 0.06  &	0.03  $\pm$	0.02 \\
    	 &    2   &       45.040 &   72.672 & 	0.26  $\pm$ 	0.08 &	0.00 	 $\pm$ 0.08  &	-0.1  $\pm$	0.09 \\
    	 &    3   &       77.472 &   99.824 & 	0.14  $\pm$ 	0.02 &	0.14 	 $\pm$ 0.02  &	0.14  $\pm$	0.02 \\
070220	 &        &       -3.512 &   27.928 & 	-0.05 $\pm$  	0.02 &	-0.06 	 $\pm$ 0.02  &	-0.05 $\pm$ 0.02 \\
070427	 &        &       -3.032 &   17.864 & 	1.62  $\pm$ 	0.4  &	0.64 	 $\pm$ 0.4   &	0.26  $\pm$	0.1  \\
070714A	 &        &       -1.176 &   8.024  &   0.34  $\pm$ 	0.2  &	0.27 	 $\pm$ 0.2 	 &   0.22 $\pm$ 0.3  \\
070911	 &    1   &        7.496 &   28.984 & 	-0.14 $\pm$  	0.08 &	0   	 $\pm$ 0.08  &	 0    $\pm$	0.02 \\
     	 &    2   &       27.768 &   62.776 & 	0.32  $\pm$ 	0.03 &	0.30 	 $\pm$ 0.03  &	0.24  $\pm$	0.06 \\
070917	 &        &       -1.128 &   9.432  & 	0.25  $\pm$ 	0.02 &	0.21 	 $\pm$ 0.02  &	0.21  $\pm$	0.02 \\
080229A	 &    1   &       -9.240 &   12.408 & 	1.07  $\pm$ 	0.31 &	0.53 	 $\pm$ 0.3   &   0.45 $\pm$ 0.2  \\
    	 &    2   &       24.744 &   52.040 & 	0.30  $\pm$ 	0.02 &	0.32 	 $\pm$ 0.02  &	0.22  $\pm$	0.03 \\
080328	 &    1   &      -12.176 &   54.848 & 	0.62  $\pm$ 	0.26 &	0.22 	 $\pm$ 0.3   &	0.16  $\pm$	0.08 \\
     	 &    2   &       65.744 &   101.824& 	0.70  $\pm$ 	0.08 &	0.45 	 $\pm$ 0.09  &	0.48  $\pm$	0.2  \\
080409	 &    1   &       -2.088 &   6.552  &   0.23  $\pm$ 	0.05 &	0.34 	 $\pm$ 0.05  &	0.35  $\pm$	0.2  \\
     	 &    2   &       5.960  &   13.336 &   0.09  $\pm$ 	0.03 &	0.00 	 $\pm$ 0.03  &	0.08  $\pm$	0.04 \\
080426	 &        &      -1.376  &   3.232  &   0.10  $\pm$ 	0.01 &	0.08 	 $\pm$ 0.02  &	0.08  $\pm$	0.02 \\
080613B	 &        &      -7.784  &   34.680 & 	0.05  $\pm$ 	0.01 &	0.03 	 $\pm$ 0.02  &	0.03  $\pm$	0.02 \\
080701	 &        &      -5.096  &   7.336  & 	1.34  $\pm$ 	0.40 &	1.49 	 $\pm$	0.4  &	1.58  $\pm$	0.5  \\
080714	 &        &      -4.000  &   6.752  & 	0.53  $\pm$ 	0.14 &	0.42 	 $\pm$ 0.1 	 &   0.16 $\pm$ 0.3  \\
080727B	 &    1   &      -0.368  &   4.928  &   -0.05 $\pm$ 	0.02 &	-0.05 	 $\pm$ 0.02  &	-0.06 $\pm$ 0.02 \\
    	 &    2   &       5.408  &   11.152 &   0.07  $\pm$ 	0.02 &	0.08 	 $\pm$ 0.02  &	0.06  $\pm$	0.02 \\
080727C	 &        &      -9.032  &   52.120 & 	0.36  $\pm$ 	0.1  &	0.34 	 $\pm$ 0.1 	 &   0.37 $\pm$ 0.2  \\
080915B	 &        &      -0.768  &   2.416  &   -0.04 $\pm$ 	0.02 &	-0.05 	 $\pm$ 0.02  &	-0.05 $\pm$ 0.03 \\
081210	 &        &       6.720  &   23.120 & 	0.05  $\pm$ 	0.07 &	0.10 	 $\pm$ 0.07  &	0.05  $\pm$	0.02 \\
090113	 &    1   &      -3.072  &   3.376  &   0.20  $\pm$    0.04  &	0.10 	 $\pm$ 0.04  &	0.19  $\pm$	0.09 \\
     	 &    2   &      3.264   &   5.328  &   0.16  $\pm$    0.05  &	0.02 	 $\pm$ 0.05  &	0.05  $\pm$	0.02 \\
     	 &    3   &      4.928   &   12.176 &   0.15  $\pm$    0.1   &	0.11 	 $\pm$ 0.1   &	0.19  $\pm$	0.1  \\
090129	 &        &      -3.472  &   23.872 & 	0.46  $\pm$ 	0.08 &	0.37 	 $\pm$ 0.08  &	0.30  $\pm$	0.1  \\
090201	 &        &      -4.256  &   3.920  &   0.12  $\pm$ 	0.2  &	0.08 	 $\pm$ 0.2 	 &   0.13 $\pm$ 0.1  \\
090201	 &        &      4.736   &   15.888 & 	0.56  $\pm$ 	0.2  &	0.50 	 $\pm$ 0.2 	 &   0.70 $\pm$ 0.3  \\
090301A	 &    1   &      -6.352  &   14.336 & 	0.32  $\pm$ 	0.08 &	0.11 	 $\pm$ 0.08  &	0.11  $\pm$	0.02 \\
     	 &    2   &      14.464  &   19.440 &   0.16  $\pm$ 	0.04 &	0.21 	 $\pm$ 0.04  &	0.22  $\pm$	0.05 \\
     	 &    3   &      21.104  &   28.784 &   0.15  $\pm$ 	0.02 &	0.11 	 $\pm$ 0.02  &	0.11  $\pm$	0.02 \\
     	 &    4   &      29.872  &   37.584 &   0.07  $\pm$ 	0.02 &	0.06 	 $\pm$ 0.02  &	0.06  $\pm$	0.02 \\
090401B	 &    1   &      -0.448  &   6.288  &   0.11  $\pm$ 	0.02 &	0.06 	 $\pm$ 0.02  &	0.06  $\pm$	0.02 \\
     	 &    2   &       6.192  &   13.104 &   0.10  $\pm$ 	0.02 &	0.03 	 $\pm$ 0.02  &	0.03  $\pm$	0.02 \\
090404	 &    1   &      14.928  &   36.528 & 	0.19  $\pm$ 	0.04 &	0.05 	 $\pm$ 0.05  &	0.11  $\pm$	0.06 \\
    	 &    2   &      33.504  &   45.600 & 	-0.21 $\pm$ 	0.1  &	-0.16 	 $\pm$ 0.1   &	0.08  $\pm$	0.1  \\
090518	 &        &      -5.888  &   3.904  &   -0.14 $\pm$ 	0.02 &	0.13 	 $\pm$ 0.03  &	0.11  $\pm$	0.03 \\
090530	 &        &      -3.024  &   7.808  & 	0.02  $\pm$ 	0.03 &	0.10 	 $\pm$ 0.03  &	0.06  $\pm$	0.03 \\
090709A	 &    1   &     -21.712 &   18.512 & 	0.27  $\pm$ 	0.05 &	0.27 	 $\pm$ 0.05  &	0.19  $\pm$	0.06 \\
    	 &    2   &     18.032  &   66.400 & 	0.09  $\pm$ 	0.02 &	0.16 	 $\pm$ 0.02  &	0.16  $\pm$	0.04 \\
090813	 &    1   &     -0.952  &   3.208  &     0.02 $\pm$ 	0.02 &	0.00 	 $\pm$ 0.02  &	0.05  $\pm$	0.02 \\
    	 &    2   &      5.384  &   9.304  &     0.16 $\pm$ 	0.1  &	0.00 	 $\pm$ 0.1   &	0.06  $\pm$	0.07 \\
090929B	 &        &      -9.064  &   53.800 & 	0.28  $\pm$ 	0.08 &	0.29 	 $\pm$ 0.08  &	0.18  $\pm$	0.06 \\
091020	 &        &      -7.136  &   23.568 & 	-0.02 $\pm$ 	0.05 &	0.00 	 $\pm$ 0.05  &	0.37  $\pm$	0.2  \\
100111A	 &        &      -5.512  &   11.752 & 	0.27  $\pm$ 	0.07 &	0.26 	 $\pm$	0.08 &	0.29  $\pm$	0.1  \\
100423A	 &        &      -3.992  &   4.552  &  0.06   $\pm$    0.02  &	0.03 	 $\pm$ 0.02  &	0.03  $\pm$	0.02 \\
100522A	 &    1   &      -2.184  &   7.752  &  -0.09  $\pm$ 	0.02 &	0.03 	 $\pm$ 0.02  &	0.03  $\pm$	0.02 \\
    	 &    2   &       23.432 &   39.672 & 	0.56  $\pm$ 	0.3  &	0.05 	 $\pm$ 0.3 	 &  0.24  $\pm$ 0.5  \\
100615A	 &    1   &     -1.60   &   9.120  & 	0.62  $\pm$ 	0.1  &	0.51 	 $\pm$ 0.1   &	0.51  $\pm$	0.08 \\
    	 &    2   &      8.976  &   22.784 & 	0.35  $\pm$ 	0.07 &	0.16 	 $\pm$ 0.07  &	0.16  $\pm$	0.04 \\
    	 &    3   &      22.496 &   48.992 & 	-0.49 $\pm$ 	0.08 &	-0.34 	 $\pm$ 0.09  &	-0.5  $\pm$	0.2  \\
100619A	 &    1   &      -7.824 &   16.592 & 	0.28  $\pm$ 	0.05 &	0.06 	 $\pm$ 0.05  &-0.02   $\pm$	0.02 \\
    	 &    2   &      79.648 &   99.792 & 	0.03  $\pm$ 	0.02 &	-0.03 	 $\pm$ 0.02  &	0.06  $\pm$	0.03 \\
100621A	 &        &      -3.288  &   42.088 & 	1.45  $\pm$ 	0.1  &	1.07 	 $\pm$ 0.13  &	1.07  $\pm$	0.1  \\
100704A	 &        &      -9.032  &   25.512 & 	1.14  $\pm$ 	0.2  &	0.78 	 $\pm$ 0.23  &	1.04  $\pm$	0.4  \\
100816A	 &        &      -1.360  &   4.176  &   0.12  $\pm$ 	0.02 &	0.10 	 $\pm$ 0.01  &	0.10  $\pm$	0.02 \\
100906A	 &    1   &      -3.248  &   21.616 & 	0.50  $\pm$ 	0.04 &	0.34 	 $\pm$ 0.05  &	0.27  $\pm$	0.05 \\
    	 &    2   &      93.024  &   131.408& 	0.72  $\pm$ 	0.2  &	0.45 	 $\pm$ 0.18  &	0.5   $\pm$ 0.2  \\
\enddata
\tablecomments{Each Column represents the same as table~\ref{tb1-1}.}
\end{deluxetable}

\end{document}